\documentclass[twocolumn,showkeys,superscriptaddress,preprintnumbers,amsmath,amssymb,showpacs,prl]{revtex4-1}
\usepackage{graphicx}
\usepackage{dcolumn}
\usepackage{bm}
\usepackage{amsmath}

\usepackage{color}
\usepackage{hyperref}
\usepackage{natbib}
\usepackage{scrextend}
\usepackage{braket}
\usepackage{soul}
\usepackage{CJK}

\newcommand{\yz}{\textcolor{black}}
\newcommand{\yznew}{\textcolor{black}}

\newcommand{\js}{\textcolor{black}}
\begin{document}
\begin{CJK*}{UTF8}{gbsn}


\title{Shear\yznew{-}jammed, fragile, and steady states in homogeneously strained granular materials}
\author{Yiqiu Zhao (赵逸秋)}
\email{yiqiu.zhao@duke.edu}
\affiliation{Department of Physics \& Center for Non-linear and Complex Systems,Duke University, Durham, NC, 27708, USA}

\author{Jonathan Bar\'{e}s}
\email{jb@jonathan-bares.eu}
\affiliation{Department of Physics \& Center for Non-linear and Complex Systems,Duke University, Durham, NC, 27708, USA}
\affiliation{Laboratoire de M\'{e}canique et G\'{e}nie Civil, Universit\'{e} de Montpellier, CNRS, Montpellier, 34090, France }

\author{Hu Zheng (郑虎)}
\email{tjzhenghu@gmail.com}
\affiliation{Department of Physics \& Center for Non-linear and Complex Systems,Duke University, Durham, NC, 27708, USA}
\affiliation{Department of Geotechnical Engineering, College of Civil Engineering, Tongji University, Shanghai, 200092, China}
\affiliation{School of Earth Science and Engineering, Hohai University, Nanjing, Jiangsu, 211100, China}

\author{Joshua~E.~S.~Socolar}
\email{socolar@phy.duke.edu}
\affiliation{Department of Physics \& Center for Non-linear and Complex Systems,Duke University, Durham, NC, 27708, USA}

\author{Robert~P.~Behringer}
\thanks{Deceased 10 July 2018.}
\affiliation{Department of Physics \& Center for Non-linear and Complex Systems,Duke University, Durham, NC, 27708, USA}
\date{\today}


\begin{abstract}
We study the jamming phase diagram of sheared granular material using a novel Couette shear set-up with multi-ring bottom. The set-up uses small basal friction forces to apply a volume-conserving linear shear with no shear band to a {granular system composed of} frictional photoelastic discs. The set-up can \yznew{generate} arbitrarily large shear strain due to its circular geometry\yznew{, and the shear direction can be reversed, allowing us to measure a feature that distinguishes shear-jammed from fragile states.} We report systematic measurements of {the stress, strain and contact network structure} at phase boundaries that have been difficult to access by traditional experimental techniques, including the yield stress curve and the jamming curve close to \js{$\phi_{SJ}\approx 0.74$}, the smallest packing fraction supporting a shear-jammed state. \yznew{We observe fragile states created under large shear strain over a range of $\phi < \phi_{SJ}$.  We also find a transition in the character of the quasi-static steady flow centered around $\phi_{SJ}$ on the yield curve as a function of packing fraction. Near $\phi_{SJ}$, the average contact number, fabric anisotropy, and non-rattler fraction all show a change of slope.  Above $\phi_{F}\approx 0.7$ the  
steady flow shows measurable deviations from the basal linear shear profile, and above $\phi_c\approx 0.78$
the flow is localized in a shear band.}
\end{abstract}

\keywords{Granular matter, Shear jamming, Strain amplitude, Couette multi-ring bottom geometry, Photoelasticity}

\maketitle

\end{CJK*}

\begin{figure}
\centering
    \includegraphics[width=0.99\columnwidth]{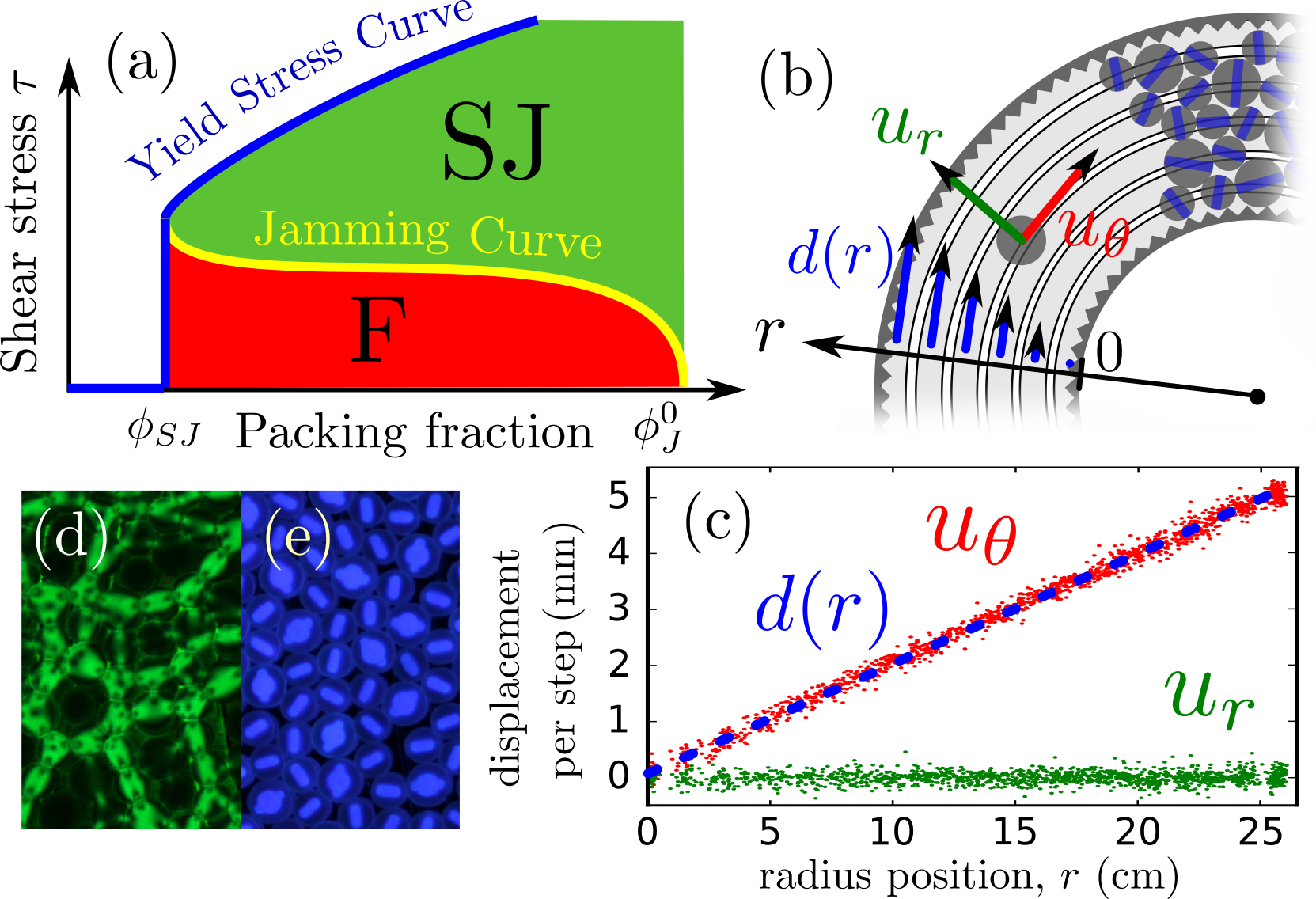}
    \caption{(color online) (a) The jamming phase diagram in the shear stress $\tau$ and packing fraction $\phi$ plane adapted from \cite{behringer2018_rpp}. Only $\phi<\phi_J^0$ part of the diagram is shown. The yield stress curve and the jamming curve are highlighted in blue and yellow, respectively. (b) Schematic of the multi-ring Couette set-up. 21 horizontal concentric rings rotate step-wisely to quasi-statically shear bidisperse photoelastic discs. For each shear step, ring at position $r$ rotates by an arc length $d(r)$. The nominal shear strain is defined as $\gamma=d(r)/r$. (c) Particle displacements in radial ($u_r$) and azimuthal ($u_{\theta}$) directions in a shear step for a dilute system ($\phi=0.57$). The dashed line is the linear basal profile $d(r)$. After each shear step the system is imaged in UV light (e) and in polarized green light (d).
    }
    \label{fig1}
\end{figure}

When a granular material prepared in a stress free state is sheared, it can make a transition into a mechanically stable state through a process known as {\em shear jamming} \cite{Bi2011_nat}. Shear jamming occurs in many different systems, including glasses \cite{Urbani2017_prl}, suspensions \cite{Han2016_natcom,majumdar2017_pre,han2018_prf,james2018_naturematerial,chen2019_nc,seto2019_gm} and dry granular matter with \cite{Howell1999_prl,Bi2011_nat,Ren2013_prl,Zheng2014_epl,dong2018_prl,otsuki2018_arxiv} or without \cite{chen2018_pre,Kumar2016_gm,Bertrand2016_pre,Baity2017_jps} friction. In 2011, Bi $et~al.$ \cite{Bi2011_nat} provided a jamming phase diagram {(Fig.~\ref{fig1}(a))} that extended the Liu-Nagel framework \cite{Liu1998_nat} by including a region of shear-jammed (SJ) states for frictional granular materials at finite shear stress with packing fractions $\phi$ between a critical value $\phi_{SJ}$ and $\phi_J^0$, the isotropic jamming packing fraction for frictionless particles. Starting from a stress free state, applying shear strain $\gamma$ can lead to two different types of jammed states: fragile (F) states that are only stable for compatible loads, and SJ states that are stable to reverse shear \cite{Cates1998_prl,Bi2011_nat}. A minimum shear strain $\gamma_{SJ}(\phi)$ is needed to create a SJ state for fixed $\phi$. In the past decade, many efforts have focused on explaining the origin of rigidity in sheared granular matter with $\phi$ close to the high packing fraction portion of the jamming curve (the yellow curve in Fig.~\ref{fig1}(a)) \cite{Bi2011_nat,Kumar2016_gm,dong2018_prl,vinutha2016_npl,sarkar2013_prl,sarkar2015_pre,sarkar2016_pre}. However, less attention 
has been paid to other parts of the phase diagram, in particular to the yield stress curve, which is important for the rheology of dense granular flow, or to the jamming curve close to the critical packing fraction $\phi_{SJ}$, where the relation between the shear strain $\gamma$ and jamming has not been experimentally determined. 

Experimental measurements of the phase boundaries in the jamming phase diagram are challenging because it is hard to create SJ states without the formation of a shear band and the associated heterogeneities in the packing fraction $\phi$ and strain field \cite{Veje1999_pre,Zhang2010_gm,Moosavi2013_prl,Coussot2002_prl,Fenistein2003_nat,Ren2013_prl}. In 2013, Ren $et~al.$ \cite{Ren2013_prl} developed a multi-slat, simple shear setup that avoids shear banding, which revealed a distinction between F and SJ states \cite{sarkar2013_prl,sarkar2016_pre}. However, their multi-slat setup had a strain limit ($\sim 60\%$) \cite{Ren2013_prl}, and thus could not access the yield stress curve or the SJ states near $\phi_{SJ}$, where $\gamma_{SJ}$ keeps growing as $\phi\rightarrow\phi_{SJ}^+$ \cite{Bi2011_nat,Bertrand2016_pre,Kumar2016_gm,Zheng2014_epl}.

In this letter we solve this challenge using a multi-ring Couette shear set-up, which applies a linear shear strain field using basal friction forces to drive the system until it becomes shear-jammed. This form of driving may be thought of as a physical implementation of the algorithm used in certain athermal, quasistatic (AQS) simulations \cite{maloney2006_pre,vinutha2016_npl}. With our apparatus, we can also keep shearing the jammed system using boundary racks to measure the yield stress curve. By shearing a layer of photoelastic disks, we for the first time \yznew{experimentally} map out the phase boundaries in the jamming phase diagram close to $\phi_{SJ}$, including the yield stress curve and the jamming curve. We find that fragile states exist below $\phi_{SJ}$ that were not included in the traditional phase diagram \cite{Bi2011_nat}. Moreover, we find two transitions on the yield stress curve: ($i$) above \yznew{$\phi_{F}\approx 0.7$}, the steady states no longer deform linearly under shear, and ($ii$) above $\phi_{c} \approx 0.78$ their deformation field becomes localized. We relate those transitions to the contact network structures.


\noindent \textit{Experiments} -- The experiments are carried out with a novel multi-ring Couette shear set-up shown in Fig.~\ref{fig1}(b), which quasi-statically and linearly shears a 2D granular medium composed of bidisperse photoelastic discs with friction coefficient $0.9$ and diameters $1.59$~cm and $1.27$~cm (denoted as $d$) \footnote{See Supplemental Material at [URL will be inserted by publisher] for (1) more details of the set-up, (2) a stress strain curve of a single particle, (3) the pressure calibration, (4) the fabric tensor calculation\yznew{,} (5) raw data for reverse shear tests \yznew{and (6) the choice of noise level}.}. The ratio of the numbers of big and small particles is $1/3$. Particles have reflective paint on their bases to enable reflective photoelasticimetry \cite{Puckett2013_prl,Zhao2017_epj,Daniels2017_rsi,zadeh2019_gm}. The total number of particles is varied from $1447$ to $2101$, which corresponds to $0.56<\phi<0.82$. The Couette set-up consists of $21$ independently controlled concentric rings. The $1.2$~cm 
wide rings rotate collectively, providing weak frictional forces to the particles sitting on them. Although essential to perform the linear shear, the magnitude of basal friction is $\sim 8$ times smaller than the typical contact forces measured in the SJ states on the jamming curve (Fig.~\ref{fig1}(a)). Particles are constrained radially by outer and inner toothed boundaries of radius {$r_{out}=35.5~cm$} and {$r_{in}=8.7~cm$}. The outer boundary rotates with the rings and the inner boundary is 
fixed.

For each experiment, a stress-free random configuration is prepared. The quasi-static linear shear is then applied in a stepwise manner. For each step, the ring at radial position $r$ rotates through an arc length $d(r)=\gamma r$. The function $d(r)$ sets the `basal profile' and $\gamma$ is called the `shear strain' by analogy with traditional simple shear \cite{Ren2013_prl}. We note that $\gamma$ is not the physical shear strain, \textit{i.e.}, the off-diagonal element of the strain tensor, $\varepsilon_{r\theta}= \partial_r  d(r)  -  d(r)/ (r+r_{in})=\gamma r_{in}/(r+r_{in})$ \cite{landau1986_book}. During a rotation step, in which $\delta \gamma = 0.6\%$, the shear rate is $\dot{\gamma} \sim 10^{-3}s^{-1}$. After each step, the rings stop for $10$~s to let the system reach a static state. As plotted in Fig.~\ref{fig1}(c), for a dilute system, the azimuthal particle displacements $u_{\theta}$ per step follow $d(r)$, and the radial displacements $u_{r}$ fluctuate around zero. No shear band is observed. \js{We apply large forward strains to measure the yield stress curve, and the strain direction is then reversed to \yznew{distinguish fragile and shear-jammed states}.}

The system is sequentially lit from the top by circular polarized green light, and from the side by ultra-violet (UV) light \cite{Note1}. Between two consecutive shear steps, after reaching a static state, the system is imaged (Canon EOS 70D, $5472 \times 3648$~px$^2$) through a circular polarizer with UV and polarized lights. UV images (Fig.~\ref{fig1}(e)) give particle positions. The polarized images (Fig.~\ref{fig1}(d)) give stress and contact information. We measure the pressure $P$, defined as the trace of the force moment tensor \cite{Bi2011_nat,Ren2013_prl}, using the averaged squared intensity gradient \cite{Howell1999_prl,Ren2013_prl,Daniels2017_rsi,zadeh2019_gm,zhao2019_njp,zhao2019_gm} of the polarized image \cite{Note1}. A sheared system must develop a non-zero $P$ to resist finite shear stress $\tau$. We also measure the non-rattler contact number $Z_{nr}$, defined as the mean contact number among stressed grains \cite{Majmudar2007_prl,Bi2011_nat,Daniels2017_rsi} \yz{(see \cite{Daniels2017_rsi} for a detailed description of the detection algorithm)}, the non-rattler fraction $f_{nr}$, defined as the number fraction of stressed grains, and the fabric anisotropy $\rho$, defined as the ratio between the difference and the sum of the eigenvalues of the fabric tensor \cite{Note1}.


\begin{figure}[!h]
    \centering
    \includegraphics[width=0.99\columnwidth]{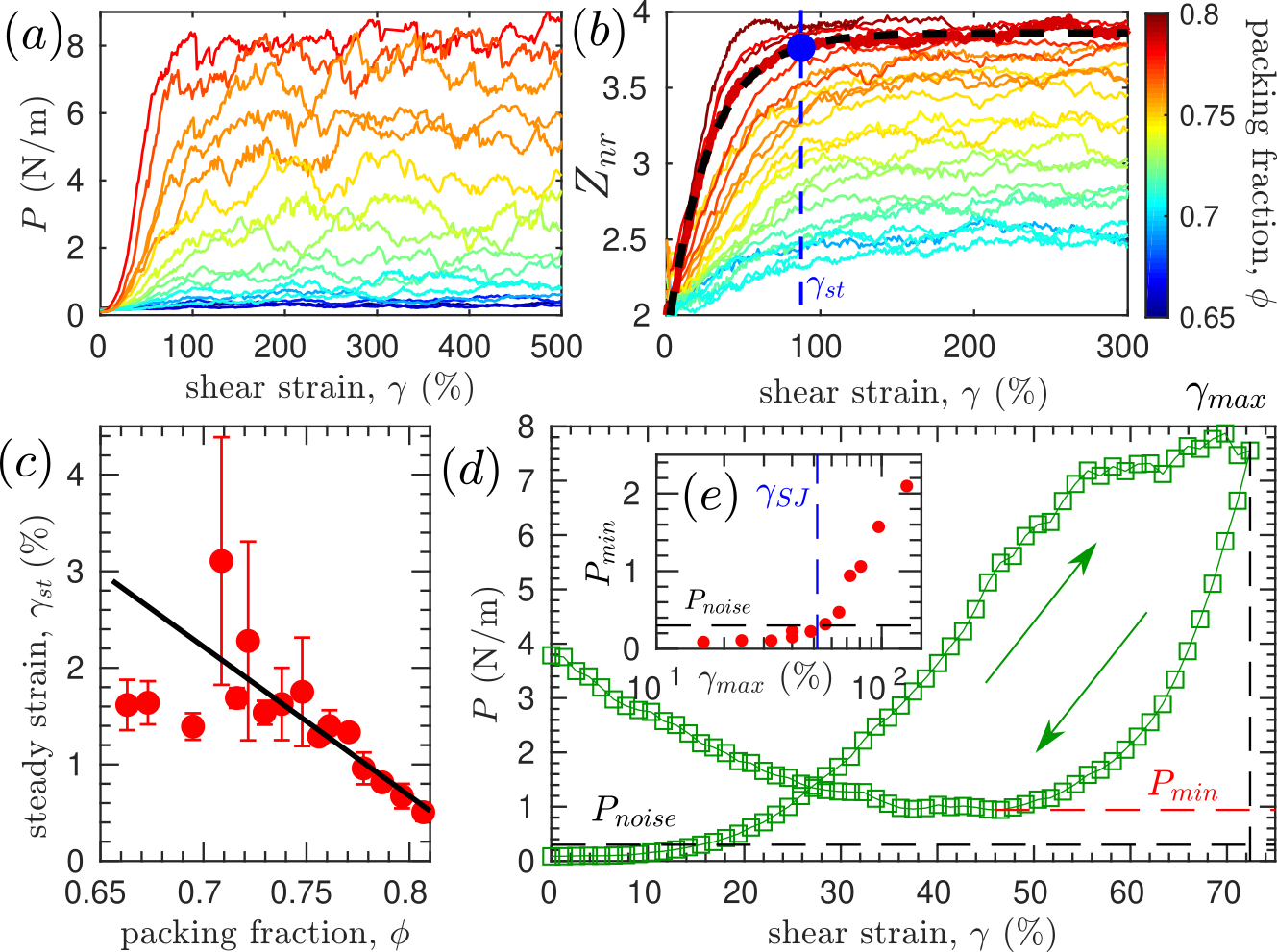}
    \caption{(color online) (a-b) The pressure $P$ and the non-rattler contact number $Z_{nr}$ versus $\gamma$ for different $\phi$ \yz{during forward shear}. The dashed black curve in (b) plots the exponential fit by Eq.~(\ref{eq_Q}) for $Z_{nr}(\gamma)$ with $\phi=0.78$. The blue dashed line shows the $\gamma_{st}$ value for this run. (c) Strain needed to reach the steady regime, $\gamma_{st}(\phi)$. The black line is a linear fit $\gamma_{st}\propto(\phi-\phi_0)$ for $\phi>0.72$.
    \yz{(d) $P$ versus $\gamma$ for a typical reverse shear test ($\phi=0.781$) with forward shear strain $\gamma_{max}$. (e) The minimum pressure, $P_{min}$, during the reverse shear versus $\gamma_{max}$ for $\phi = 0.781$. $\gamma_{SJ}$ is the minimum $\gamma_{max}$ for which $P_{min}>P_{noise}=0.3$~N/m \yznew{\cite{Note1}}.} }
    \label{fig2}
\end{figure}

\noindent \textit{Results} --  Figures~\ref{fig2}(a) and (b) show pressure $P$ and non-rattler contact number $Z_{nr}$ versus shear strain $\gamma$, for typical runs with different $\phi$. For a given $\phi$, after a transient growth regime, both $Z_{nr}$ and $P$ fluctuate around constant values that define the yield stress curve. We refer the associated stress as the ``steady states'' stress. We find that $Z_{nr}$ 
can be fitted to:
\begin{equation}
	Q=Q_{st}+c* e^{-\gamma/\gamma_c}
\label{eq_Q}
\end{equation}
where $Q$ can be $Z_{nr}$, $f_{nr}$ or $1-\rho$, and $Q_{st}$, $c$ and $\gamma_c$ are fit parameters. An example fit for $Z_{nr}(\gamma)$ with $\phi=0.76$ is plotted in Fig.~\ref{fig1}(b). We find that the steady regime \yz{has been}
reached at $\gamma_{st} \equiv 3 \gamma_c$ for all state variables, where $\gamma_c$ is obtained from the fits for $Z_{nr}$. Figure~\ref{fig2}(c) shows $\gamma_{st}(\phi)$, where a linear fit $\gamma_{st}\propto (\phi-\phi_0)$ for $\phi>0.72$ gives $\phi_0=0.84 \pm 0.02$, close to the frictionless isotropic jamming density \cite{OHern2003_pre}. \js{The slope is $-1545\pm 427$ (\%).}

\begin{figure}
    \centering
    \includegraphics[width=0.99\columnwidth]{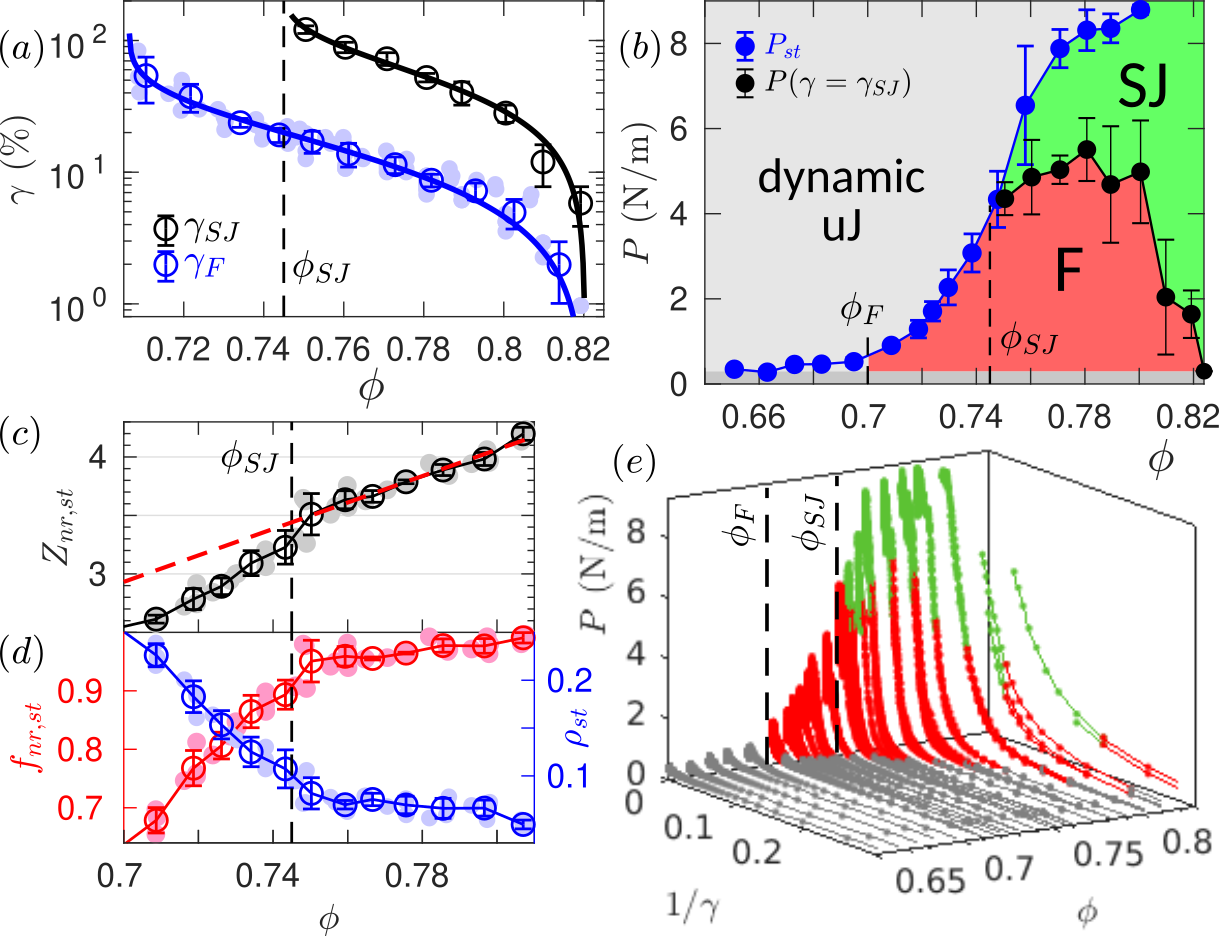}
    \caption{(color online) \yz{(a)} Strain needed to create fragile state, $\gamma_F$ (blue), and SJ state, $\gamma_{SJ}$ (black). The minimum packing fraction for \yz{SJ}
    states is  \yz{$\phi_{SJ}\approx 0.745$}. The blue \yz{and black} solid curves \yz{are fitted with Eq.~(\ref{eq_gamS}). In (a), (c) and (d), solid gray circles are raw data and open circles are averaged data.}
    \yz{(b)} The jamming phase diagram in the $(P,\phi)$ plane built from our data. The dynamic unjammed (uJ), fragile (F) and shear jammed (SJ) states are separated by the yield stress curve $P_{st}$ and the jamming curve, $P(\gamma=\gamma_{SJ})$. 
    The dark gray region below the {noise level 0.3~N/m} indicates {static} unjammed states. \yz{$\phi_F$ is the minimum packing fraction for fragile states.}  (c-d) The \yz{steady state} non-rattler contact number $Z_{nr,st}$, \yz{non-rattler fraction $f_{nr,st}$ and fabric anisotropy $\rho_{st}$ \yz{obtained from Eq.~(\ref{eq_Q})}. Note the change in slope near $\phi_{SJ}$ in all three cases. The red dashed line in (c) shows a linear fit using data above $\phi_{SJ}$.} 
    (e) Surface plot of all static states measured \yz{during forward shear experiments} in the space of $P$, $\phi$ and inverted strain $1/\gamma$ space. Smooth curves join states accessed in a single run. States are labeled using same color code as in (b). 
    }
    \label{fig3}
\end{figure}

\js{We identify a system as} shear-jammed (SJ) \js{if under reverse shear the pressure never drops below the noise threshold $P_{noise}=0.3$~N/m \yznew{\cite{Note1}}, which indicates that the system resists the reversed stress rather than simply allowing a reversion to a stress-free (unjammed) state.} \yz{Figure \ref{fig2}(d) shows the evolution of $P$ during a shear cycle for a system with $\phi=0.781$.  Figure~\ref{fig2}(e) plots the dependence of the minimum pressure $P_{min}$ during reverse shear on the maximum forward shear strain $\gamma_{max}$, from which we extract the minimum strain, $\gamma_{SJ}$,}
required to create a SJ state. \yz{We find no SJ state for $\phi=0.74$  even when $\gamma_{max}\gg\gamma_{st}$ \cite{Note1}. For $\phi=0.75$ $\gamma_{SJ}\approx \gamma_{st}$. The minimum packing fraction that supports shear jamming must lie between these two values:  $\phi_{SJ}=0.745\pm 0.005$.} 
Figure \ref{fig3}(a) plots the relation between $\gamma_{SJ}$ and $\phi$,
which can be fitted using \yz{a form suggested by}
\cite{Kumar2016_gm},
\begin{equation}
	\gamma_{SJ}(\phi)=\gamma_{b} \left[\ln\left(\frac{\phi_{J}^0-\phi_{SJ}}{\phi-\phi_{SJ}}\right)\right]^{\alpha}
\label{eq_gamS}
\end{equation}
\noindent where \yz{$\phi_{SJ}=0.745$ is preset and }the fit parameters are \yz{
$\alpha = 0.68\pm0.11$, $\gamma_{b}=64\pm 6(\%)$ and $\phi_{J}^0=0.820\pm0.005$.}

\yznew{In this work,} fragile (F) states 
\yznew{refer to} states with non-zero pressure ($P>P_{noise}$) \yz{
and have $P_{min}<P_{noise}$ at some point in the reverse shear process}. As shown in \js{Fig.~\ref{fig3}(a)}, we find
\yz{$\gamma_F$}, the minimum strain required to create a fragile state, \yz{also follows Eq.~\ref{eq_gamS}. In this fit, we take $\phi_J^0=0.82$ from the previous fit, and we determine $\phi_F$, the minimum packing fraction for fragile states, from the fit, obtaining $\phi_F = 0.706\pm 0.003$ along with $\gamma_b = 19\pm 2$~(\%) and $\alpha = 0.86\pm0.12$.}
\yznew{We also note, however, that the divergence predicted by Eq.~\ref{eq_gamS} near $\phi_{SJ}$ and $\phi_{F}$ is not clearly seen in our data.} 
Below $\phi_F$, the steady state pressure falls \yz{to a plateau value near} the noise level. 

Figure~\ref{fig3}(b) shows the experimentally constructed jamming phase diagram in the $(P,\phi)$ space. The yield stress curve is the $P_{st}(\phi)$ curve, showing the average steady state pressure for each $\phi$. $P_{st}$ \js{increases monotonically from $\phi_F$ and appears to have an inflection point at $\phi_{SJ}$. However, for the steady states above $\phi\approx 0.78$, the pressure of some particles becomes so large that their photoelastic fringes can not be resolved, likely leading to artificially low pressure measurements.}
$P_{st}(\phi)$ also separates SJ states and the dynamic unjammed states, which have non-zero shear rates. The jamming curve is also plotted as the $P(\phi,\gamma_{SJ})$ curve, which consists of the pressure value for each $\phi$ at the jamming strain $\gamma_{SJ}$. The gray region below $P_{noise}$ refers to the static unjammed states without measurable stress. Figure~\ref{fig3}(e) extends (b) by including the inverted strain axis and plots all the static states measured \yz{during the forward shear process} in the $(P,\phi,1/\gamma)$ space, highlighting their dependence on the driving strain $\gamma$. \yz{A state is labeled SJ when the shear strain exceeds $\gamma_{SJ}$ determined using Eq.~\ref{eq_gamS}.} All static SJ (green), F (red) and unjammed (gray) states lie approximately on a smooth surface in the 3D space.

\begin{figure}
    \centering
    \includegraphics[width=0.99\columnwidth]{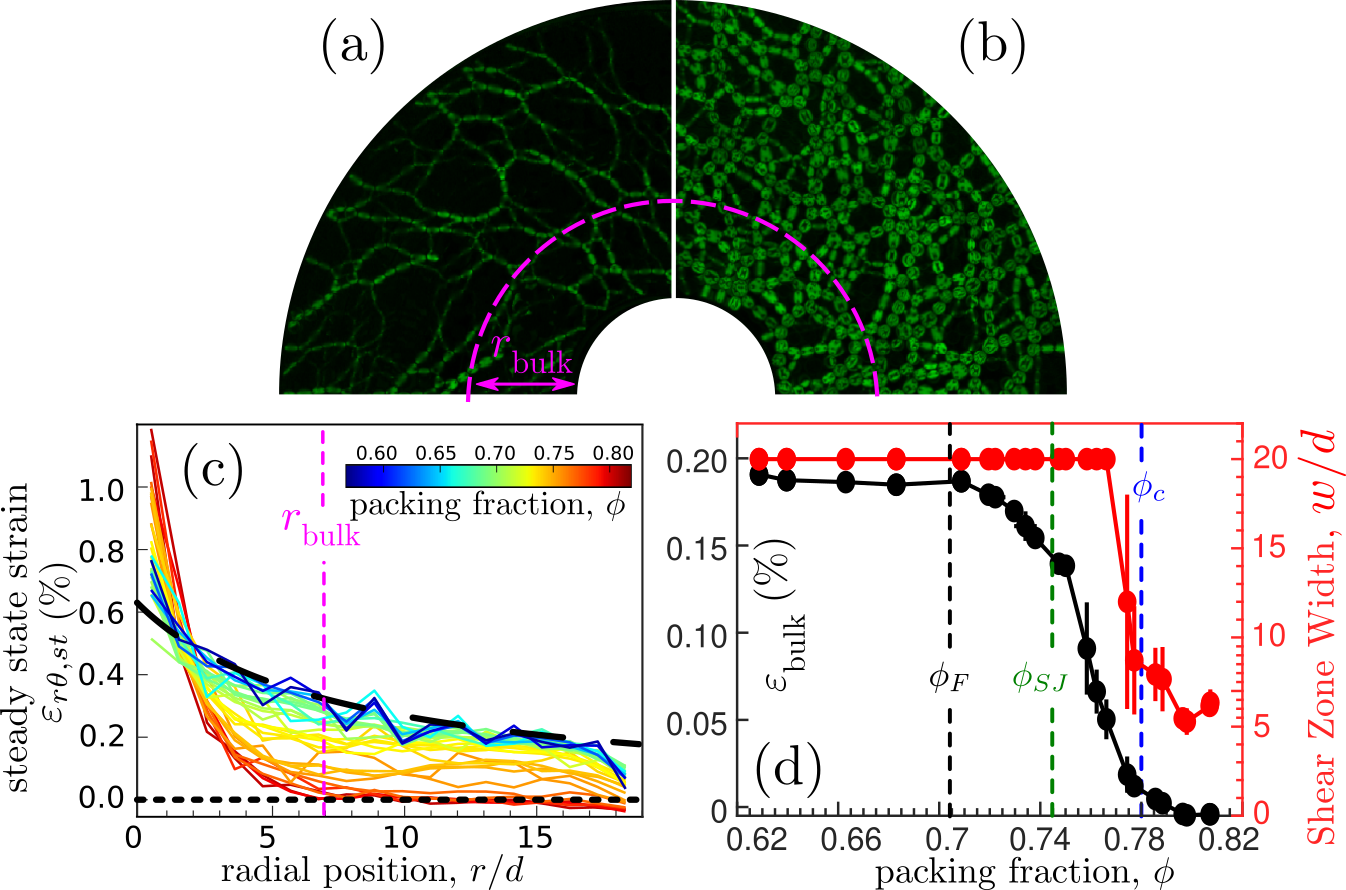}
    \caption{(color online) {(a-b) Polarized images showing force networks of typical steady states with packing fraction $0.72$ and $0.78$.}
    (c) Physical shear strain per step {averaged over steady} states, $\varepsilon_{r\theta,st}(r)$ for different $\phi$, labeled by the colorbar. The black dashed curve shows the basal profile {with linear nominal strain $\gamma$}. $r_{\text{bulk}}=7d$ is where $\varepsilon_{r\theta,st}$ vanishes for $\phi>\phi_c$. 
    (d) $\varepsilon_{\text{bulk}}$, defined as the averaged $\varepsilon_{r\theta,st}$ for $r>r_{\text{bulk}}$, drops at $\phi_{SJ}$ and vanishes at $\phi_c$.  {Same figure plots the width of shear zone $w$, defined as the range of $r$ that $\epsilon_{r\theta,st}$ is non-zero. $w(\phi)$ drops to $r_{bluk}$ at $\phi\approx\phi_c$.}}
    \label{fig4}
\end{figure}

To quantify the contact network structure on the yield stress curve, we measure $Z_{nr,st},f_{nr,st}$ and $\rho_{st}$, which are obtained from fits to the form of Eq.~(\ref{eq_Q}). Figures~\ref{fig3}(c) and (d) show a change in slope in \yz{all three state variables at a packing fraction slightly above $\phi_{SJ}$. The red dashed line in Fig.~\ref{fig3}(c) is the linear fit using data with $\phi>\phi_{SJ}$, which highlights the change in behavior at $\phi_{SJ}$.} 
Figures~\ref{fig4}(a) and (b) show two polarized images taken from the steady regime with packing fractions $0.72$ and $0.78$\yz{, showing typical force network in F and SJ states.}


When the system is shear-jammed, the basal friction becomes unimportant, and the particle displacement field deviates from the basal profile. Based on the azimuthal displacement field per shear step averaged over the steady states, $u_{\theta,st}(r)$, we calculate the off-diagonal element of the strain tensor $\varepsilon_{r\theta,st}(r) = \partial_r  u_{\theta,st}(r)  -  u_{\theta,st}(r)/ (r+r_{in})$ \cite{landau1986_book}, which gives the mean physical shear strain field for steady states (Fig.~\ref{fig4}(c)). We also measure the width of the shear zone $w$, which is the $r$ value beyond which $\varepsilon_{r\theta,st}$ becomes {smaller than the noise level 0.02\%}. Figure~\ref{fig4}(d) shows $w(\phi)$ (in red), which jumps discontinuously near $\phi_c\approx 0.78$, below which $w=r_{out}-r_{in}\approx 20d$. Above $\phi_c$, $w\approx 7d$ , denoted $r_{bulk}$ in Fig.~\ref{fig4}(c). \yz{The local packing fraction in this shear band is also smaller} \js{than the global value}. The part of the system with $r>r_{bulk}$ just rotates as a solid with the moving outer boundary in the steady states for $\phi>\phi_c$. We also calculate $\varepsilon_{bulk}$, which is the averaged $\varepsilon_{r\theta,st}$ for $r>r_{bulk}$. Figure~\ref{fig4}(d) shows $\varepsilon_{bulk}$ starts to drop at \yznew{$\phi_{F}$} and becomes zero near $\phi_c$.


\noindent \textit{Concluding discussion--} We set up a multi-ring Couette device that uses small basal friction to \yznew{drive a 2D granular medium in a way that maintains a linear shear strain profile until the system becomes jammed, allowing us to probe the jamming transition close to $\phi_{SJ}$. The set-up subsequently shears the jammed system using the boundary racks, allowing a study of the yield stress curve for a wide range of packing fractions. Finally, reversing the direction of the drive allows us to distinguish shear-jammed (SJ) from fragile (F) states.}



We systematically measured the phase boundaries in the jamming phase diagram, including close to $\phi_{SJ}$, leading to the following key observations: {($i$) In our system \yz{$\phi_{SJ}\approx 0.75$}, whose value may depend on the friction coefficient $\mu$, polydispersity, and particle shape, though we expect the qualitative features of the jamming phase diagram to be the same.} ($ii$) The \yz{SJ} strain $\gamma_{SJ}$ \yz{\yznew{is well fit by}
a stretched logarithmic function of $\phi$.} \yz{The measured exponent $\alpha=0.68\pm 0.11$ is in quantitative agreement with the exponent $\alpha=1/1.37\approx 0.73$ measured from simulation of sheared 3d frictionless soft spheres \cite{Kumar2016_gm}. The same form, but with $\alpha=1$, has also been observed in experiments on shear-thickening suspensions \cite{han2018_prf}.} 
 ($iii$) We observe fragile states below  $\phi_{SJ}$, 
 which are not included in the traditional phase diagram \cite{Bi2011_nat}. In our system, small basal friction forces and particle deformability may be crucial for stabilizing the fragile force network.  ($iv$) On the yield stress curve, for increasing packing fraction, \yz{we find that $P_{st}$ has an inflection point at $\phi_{SJ}$ and that $Z_{nr,st}$, $\rho_{st}$ and $f_{nr,st}$ all show a change of slope near $\phi_{SJ}$, suggesting a physical transition in the nature of the steady states.}

\yznew{We also find that the quasi-static steady flow field changes from the non-localized basal profile for systems with $\phi<\phi_F\approx 0.7$ to a localized shear band for $\phi > \phi_c\approx 0.78$, where $\phi_F<\phi_{SJ}<\phi_c$.} 
\yznew{The} coexistence of a solid and fluid phase in slowly sheared dense granular matter has been reported in many systems \cite{Coussot2002_prl,Debregeas2001_prl,Varnik2003_prl,Losert2000_prl,Schall2010_arfm,Moosavi2013_prl,Veje1999_pre}. In this work we characterize the \yznew{contact}
network associated with the different quasistatic steady flow regimes. 
When $\phi=\phi_c$, the steady states have $\rho_{st}\approx 0.05$ and $f_{nr,st}\approx 1$, showing a \js{nearly} isotropic\yz{, fully percolated} contact network. Notably, $\phi_c\approx\phi_J^{\mu}$ with $\mu\approx 0.9$, where $\phi_J^{\mu}$ is the isotropic jamming packing fraction with friction coefficient $\mu$ \cite{Silbert2010_sm}. We also note that \yz{$Z_{nr,st}(\phi_{SJ})\approx 3.4$, similar to the mean contact number observed when a strong force network percolates in both principal directions in biaxial experiments \cite{Bi2011_nat},} and $Z_{nr,st}(\phi_c)\approx 3.9$, \yz{close to} the isostatic value for ideal 
frictionless disks \cite{Hecke2009_jpcm,OHern2003_pre}.

The results suggest several directions for further study. 
\yz{First}, our shear device can generate other basal profiles \cite{Zhao2017_epj} to study how shear jamming affects the granular rheology for shear fields found in real world applications. \yz{Second}, the set-up can create a controlled shear band, providing a new technique to study the generation and evolution of shear bands in dense granular flow.

\begin{acknowledgments}
We thank Bulbul Charkraborty, Dong Wang, \yznew{Mark D. Shattuck, Karen E. Daniels}, Dapeng Bi, Hisao Hayakawa and Michael Rubinstein for fruitful discussions. Special thanks to Dong Wang, Yuchen Zhao, Bernie Jelinek and Richard Nappi for technical support. 
This work was supported by NSF-DMR1206351, NSF-DMS1248071, and NASA NNX15AD38G. H.Z. also received support from NSFC 41672256 and NSFC(Jiangsu) BK20180074.
\end{acknowledgments}

\bibliography{main}

\appendix*
\section{Supplemental Material}

\noindent \textbf{Details of the experimental set-up} \\

\begin{figure}[!h]
\centering
\includegraphics[width=1\columnwidth]{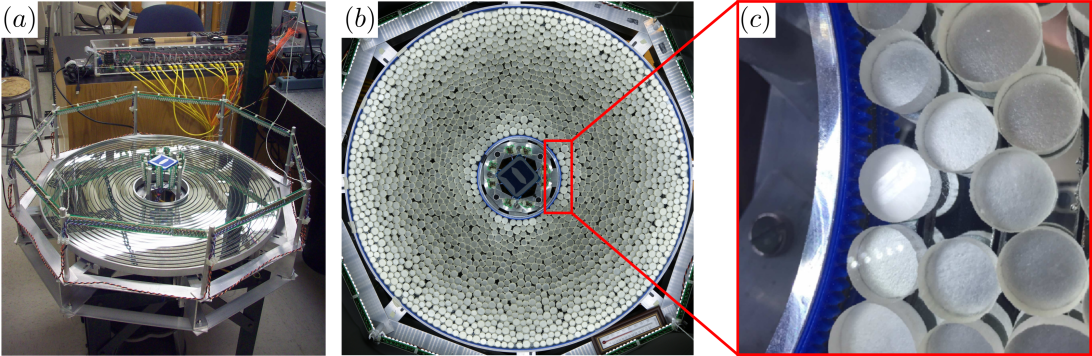}
\caption{\label{FigSetUpPicture} {(a)} Full view of the Couette shear set-up with the 21 driving glass rings and the controller in the back. {(b) Top view of the set-up with particles. (c)} Close picture of the inner wall of the Couette cell. It is covered with a blue plastic rack gear.}
\end{figure}

Figure~\ref{FigSetUpPicture}(a) shows a detailed view of the experimental set-up. {Figure~\ref{FigSetUpPicture}(b) shows a top view of the set-up with particles. }
Figure~\ref{FigSetUpPicture}(c) shows a detailed view of the rack gear fixed around the inner wall of the Couette cell. A similar rack is also fixed inside the outer wall. The racks allow for strong friction forces at the inner and outer boundaries, which in enable boundary driven shear to occur at pressures high enough that basal friction alone is not effective.

\noindent \textbf{{Details of the} imaging {system}} \\

\begin{figure}[!h]
\centering
\includegraphics[width=1\columnwidth]{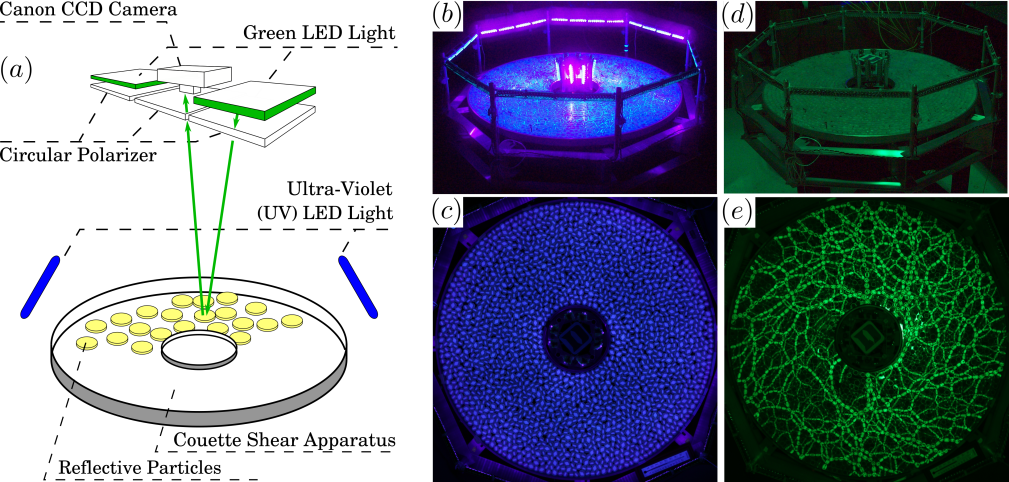}
\caption{\label{FigCameraPicture} (a) Schematic view of the imaging and lighting elements of the experimental device. The bottom of the shear apparatus as well as the bottom of particles are reflective. Some example paths for the green light are plotted. (b) Couette shear set-up with UV lighting turned on. (c) Experimental UV image revealing particle positions and orientations. (d) Couette shear set-up with green polarized light turned on. {(e) Experimental polarized image revealing stress and contact information.}}
\end{figure}

Figure~\ref{FigCameraPicture}(a) presents the imaging system used to measure stresses on individual particles~\cite{Zhao2017_epj}. This imaging system is similar to the one described in Ref.~\cite{Puckett2013_prl}. The particles are lit from the top by green flat lights laying just next to the camera. A first polarizer is set between these lights and the particles. The polarized light passes through the particles. This light, after going through each particle, is reflected by the reflective paint at the bottom of the particles.The reflected light passes again through the particles, passes through a second polarizer with the same polarization direction as the first, and is imaged by a camera. For each shear step the system is lit from the edges and the center of the rings by ultra-violet (UV) light and then from the top by polarized green light and imaged by a camera with a crossed polarizer. Figure~\ref{FigCameraPicture}(c) and~(e) show example UV and polarized images. Figure~\ref{FigCameraPicture}(b) and (d) show side views of the shear set-up under UV and green polarized light.

~\newline 

\noindent \textbf{Contact force law and the pressure measurement calibration} \\

\begin{figure}[!h]
    \centering
    \includegraphics[width=1\columnwidth]{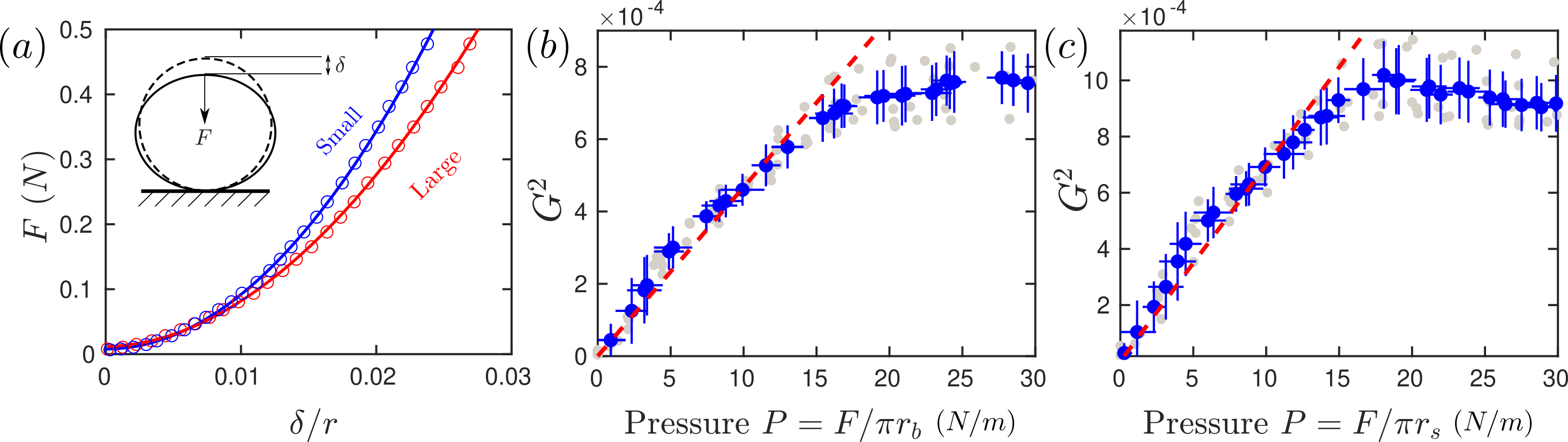}
    \caption{(a) The relation between the normal force magnitude and relative deformation, $\delta / r$, where $\delta$ is the deformation of particle along a diameter and $r$ is the particle radius. The blue solid curve is a fit to the form $F=A_s(\frac{\delta}{r})^{\alpha_s}+c_s$, giving $A_s=820\pm 88$, $\alpha_s=1.99\pm 0.03$ and $c_s=0.007\pm 0.002$. Red solid curve is a fit with $F=A_b(\frac{\delta}{r})^{\alpha_b}+c_b$ giving  $A_b=447\pm 60$, $\alpha_b=1.90\pm 0.04$ and $c_b=0.011\pm 0.003$. The insert schematic shows the diametric loading test, where the particle is pushed using a force $F$ oriented normal to a still wall. (b-c) Pressure calibration curves for large (b) and small (c) particles. The calibration is performed using the diametric test in (a). Gray and blue dots show raw and averaged data. The linear regime ($P<15N/m$) is fitted to a linear function $G^2=\frac{1}{k}P$ (dashed red line) giving $k=(2.11\pm 0.08)\times 10^4$ for large disks and $k=(1.40\pm 0.07)\times 10^4$ for small disks.}
    \label{FigCalibration}
\end{figure}

Figure~\ref{FigCalibration}(a) plots the relation between normal contact force magnitude and particle deformation. $\delta/r$ is the ratio of the change of diameter under force $F$ divided by the original particle radius. In this work, we can measure only $F<0.5N$, leading to less than 3\% relative deformation $\delta/r$. Values of $F$ are measured with respect to a measured noise threshold. For forces in this regime, $F$ is proportional to the square of the deformation: $F\propto (\delta/r)^{\alpha}$, with $\alpha \approx 2$ for large and small particles. Details of the fits can be found in the caption of Fig.~\ref{FigCalibration}.

The pressure $P$, as defined by the trace of the force moment tensor, is equivalent to the averaged pressure on each particle. For a single particle, $P=\sum_{i=1}^zF_i/2\pi r$, where $z$ is the number of contacts, $F_i$ is the normal component of the $ith$ contact force and $r$ is the particle radius. We measure $P$ using a previously developed ``$G^2$'' technique~\cite{Howell1999_prl,geng2001_prl,Daniels2017_rsi}
. $G^2$ is the sum of the squared gradient of the intensity over the pixels inside the particle. We calibrate the relation between $P$ and $G^2$ using the diametric test shown in Fig.~\ref{FigCalibration}(a). As shown in Fig.~\ref{FigCalibration}(b-c), for both small and large disks, $P=kG^2$ when $P<15N/m$, which corresponds to $F<0.5N$. For larger forces, the photoelastic fringes cannot be resolved well with the current imaging system. The values of $k$ can be found in the caption of Fig.~\ref{FigCalibration}.  As mentioned in the main text, this limit is reached at the yield stress curve for $\phi>0.77$, but does not affect other results.

~\newline

\noindent\textbf{Calculation of the fabric anisotropy}\\

In this work, the fabric tensor is calculated in order to better mimic a simple shear system. Let $S$ denote the area of the container, and let $\vec{r}_{ij}=r_{ij,x}\hat{e}_t+r_{ij,y}\hat{e}_r$ be the branch vector pointing from the center of $i$th particle to the contact between $i$th and $j$th particles. $\hat{e}_t$ and $\hat{e}_r$ are the unit tangential and radial vectors in a system centered on the $i$th particle. The fabric tensor is defined as
\begin{equation}
    \hat{R}=\frac{1}{S}\sum_{i,j} \vec{r}_{ij}\otimes \vec{r}_{ij}=\frac{1}{S}\sum_{i,j} \begin{pmatrix} 
r_{ij,x}r_{ij,x} & r_{ij,x}r_{ij,y} \\
r_{ij,y}r_{ij,x} & r_{ij,y}r_{ij,y} 
\end{pmatrix}
\end{equation}
\noindent where the summation is over all contacting pairs $(i,j)$. The fabric anisotropy is $\rho=(R_1-R_2)/(R_1+R_2)$, where $R_1$ and $R_2$ are eigenvalues of $\hat{R}$. 
~\newline

\noindent\textbf{\yz{Raw data of the reverse shear tests}}\\

\begin{figure}[!h]
\centering
\includegraphics[width = 9cm]{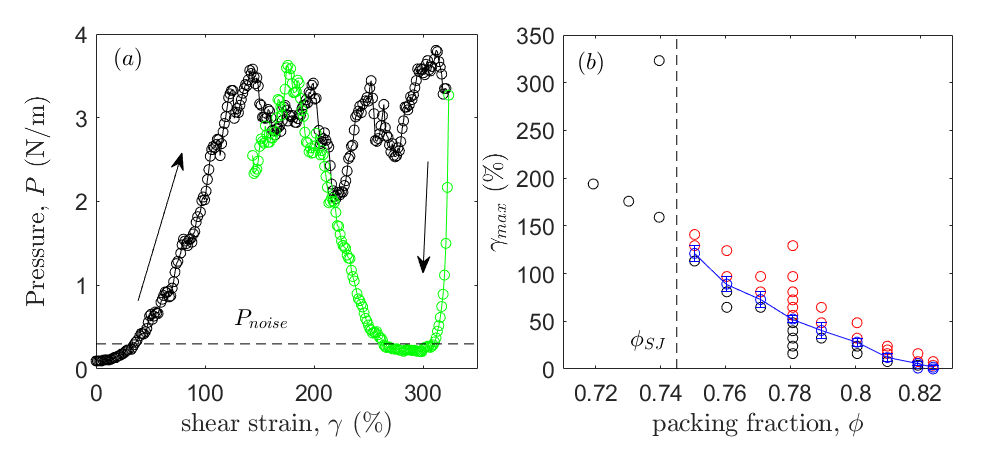}
\caption{\yz{(a) The pressure $P(\gamma)$ during the forward (black) and reverse (green) shear with $\gamma_{max}=323\%$ for packing fraction $\phi=0.74$, (b) The map of all experiments we performed to measure the jamming curve. Each black or red dot represents one reverse shear experiment with the corresponding $\phi$ and the forward shear strain amplitude $\gamma_{max}$. A black dot means $P_{min}<P_{noise}$ during the reverse shear, and a red dot means for this run $P_{min}>P_{noise}$. The blue dots show the jamming strain $\gamma_{SJ}$, defined as the strain halfway between the minimum $\gamma_{max}$ for red dots (denoted as $\gamma_{red}$) and the maximum $\gamma_{max}$ for the black dots (denoted as $\gamma_{black}$) with same $\phi$. The difference between $\gamma_{red}$ and $\gamma_{black}$ determines the error bar.} \label{jam_thresh} }
\end{figure}

 \yz{Figure~\ref{jam_thresh}(a) shows a typical run at the largest packing fraction where no shear-jammed state is observed. Even though the system has reached the steady flow regime and supports a nonzero pressure, it remains fragile.} 
 
\yz{Note that for small strains, where $P<P_{noise}$, $P$ increases slowly due to the growth of weak force chain segments that do not connect the inner and outer boundaries of the system.  Basal friction and imperfections in the ring alignments are two sources of this effect. The value of $P_{noise}$ is chosen empirically by examining photoelastic images to make sure that at least one force chain connects the inner to the outer boundary for $P>P_{noise}$. A small change in the choice of $P_{noise}$ does not alter the qualitative features of the measured phase boundaries.}

\end{document}